
\input epsf
\newif\iftwoside
\newif\ifdoublepage
\doublepagefalse
\twosidefalse
%
\hoffset=0cm
\voffset=0cm
\hsize=16.0 true cm
\vsize=23.0 true cm
\parskip=\medskipamount
%
\font\mfiverm=cmr12 scaled\magstep5
\font\mfourrm=cmr12 scaled\magstep4
\font\mthreerm=cmr12 scaled\magstep3
\font\mfivesy=cmsy10 scaled 3583
\font\mfoursy=cmsy10 scaled 2986
\font\mthreesy=cmsy10 scaled 2488
\font\mfivebf=cmbx12 scaled 3583
\font\mfourbf=cmbx12 scaled 2986
\font\mthreebf=cmbx12 scaled 2488
\font\mfiveex=cmex10 scaled 3583
\font\mfourex=cmex10 scaled 2986
\font\mthreeex=cmex10 scaled 2488
\font\mfivei=cmmi12 scaled 3583
\font\mfouri=cmmi12 scaled 2986
\font\mthreei=cmmi12 scaled 2488





 \font\twelverm=cmr12     \font\twelvei=cmmi12
 \font\twelvesy=cmsy10 scaled 1200  \font\twelveex=cmex10 scaled 1200
 \font\twelvebf=cmbx12   \font\twelvesl=cmsl12
 \font\twelvett=cmtt12  \font\twelveit=cmti12

 \skewchar\twelvei='177   \skewchar\twelvesy='60


 \def\bigfont{\spaceskip=10 pt
   \normalbaselineskip=35pt
   \abovedisplayskip 12.4pt plus 3pt minus 9pt
   \belowdisplayskip 12.4pt plus 3pt minus 9pt
   \abovedisplayshortskip 0pt plus 3pt
   \belowdisplayshortskip 7.2pt plus 3pt minus 4pt
   \smallskipamount=3.6pt plus1.2pt minus1.2pt
   \medskipamount=7.2pt plus2.4pt minus2.4pt
   \bigskipamount=14.4pt plus4.8pt minus4.8pt
   \def\rm{\fam0\mfiverm}          \def\it{\fam\itfam\mfivei}%
   \def\sl{\fam\slfam\mfivesl}     \def\bf{\fam\bffam\mfivebf}%
   \def\mit{\fam 1}                 \def\cal{\fam 2}%
   \def\tt{\fourteentt}
   \textfont0=\mfiverm   \scriptfont0=\mfourrm   \scriptscriptfont0=\mthreerm
   \textfont1=\mfivei    \scriptfont1=\mfouri    \scriptscriptfont1=\mthreei
   \textfont2=\mfivesy   \scriptfont2=\mfoursy   \scriptscriptfont2=\mthreesy
   \textfont3=\mfiveex   \scriptfont3=\mfourex  \scriptscriptfont3=\mthreeex
   \textfont\itfam=\mfivei
   \textfont\slfam=\twelvesl
   \textfont\bffam=\mfivebf \scriptfont\bffam=\mfourbf
   \scriptscriptfont\bffam=\mthreebf
   \normalbaselines\rm}

 \def\twelvepoint{\normalbaselineskip=12.4pt
   \abovedisplayskip 12.4pt plus 3pt minus 9pt
   \belowdisplayskip 12.4pt plus 3pt minus 9pt
   \abovedisplayshortskip 0pt plus 3pt
   \belowdisplayshortskip 7.2pt plus 3pt minus 4pt
   \smallskipamount=3.6pt plus1.2pt minus1.2pt
   \medskipamount=7.2pt plus2.4pt minus2.4pt
   \bigskipamount=14.4pt plus4.8pt minus4.8pt
   \def\rm{\fam0\twelverm}          \def\it{\fam\itfam\twelveit}%
   \def\sl{\fam\slfam\twelvesl}     \def\bf{\fam\bffam\twelvebf}%
   \def\mit{\fam 1}                 \def\cal{\fam 2}%
   \def\tt{\twelvett}
   \textfont0=\twelverm   \scriptfont0=\tenrm   \scriptscriptfont0=\sevenrm
   \textfont1=\twelvei    \scriptfont1=\teni    \scriptscriptfont1=\seveni
   \textfont2=\twelvesy   \scriptfont2=\tensy   \scriptscriptfont2=\sevensy
   \textfont3=\twelveex   \scriptfont3=\twelveex  \scriptscriptfont3=\twelveex
   \textfont\itfam=\twelveit
   \textfont\slfam=\twelvesl
   \textfont\bffam=\twelvebf \scriptfont\bffam=\tenbf
   \scriptscriptfont\bffam=\sevenbf
   \normalbaselines\rm}

 \def\beginlinemode{\endmode
   \begingroup\parskip=0pt \obeylines\def\\{\par}\def\endmode{\par\endgroup}}
 \def\beginparmode{\endmode
   \begingroup \def\endmode{\par\endgroup}}
 \let\endmode=\par
 {\obeylines\gdef\
 {}}
 \def\singlespace{\baselineskip=\normalbaselineskip}
 
 \def\oneandahalfspace{\baselineskip=\normalbaselineskip
   \multiply\baselineskip by 3 \divide\baselineskip by 2}
 \def\doublespace{\baselineskip=\normalbaselineskip \multiply\baselineskip by
2}
 
 \let\rawfootnote=\footnote              
 \def\footnote#1#2{{\rm\singlespace\parindent=0pt\parskip=0pt
   \rawfootnote{#1}{#2\hfill\vrule height 0pt depth 6pt width 0pt}}}
 \def\raggedcenter{\leftskip=4em plus 12em \rightskip=\leftskip
   \parindent=0pt \parfillskip=0pt \spaceskip=.3333em \xspaceskip=.5em
   \pretolerance=9999 \tolerance=9999
   \hyphenpenalty=9999 \exhyphenpenalty=9999 }

 \def\\{\cr}
 \twelvepoint            
 \oneandahalfspace         

 \overfullrule=0pt       

 \def\title                      
   {\null\vskip 3pt plus 0.2fill        \beginlinemode
    \doublespace \myraggedcenter \bigfont}

 \def\author                     
   {\vskip 3pt plus 0.2fill \beginlinemode
    \singlespace \raggedcenter}

 \def\affil                      
   {\vskip 3pt plus 0.1fill \beginlinemode
    \oneandahalfspace \raggedcenter \sl}

 \def\abstract                   
   {\vskip 3pt plus 0.3fill \beginparmode
    \oneandahalfspace
    \centerline{\bf Abstract}

                                 }

 \def\endtopmatter               
   {\endpage                     
    \body}

 \def\body                       
   {\beginparmode}               

 \def\beneathrel#1\under#2{\mathrel{\mathop{#2}\limits_{#1}}}

 \def\refto#1{$[#1]$}           

 \gdef\refis#1{\item{#1.\ }}                     

 \def\figurecaptions             
   {\endpage
    \beginparmode
    \head{Figure Captions}
 }

 \def\endpage                    
   {\vfill\eject}

 \def\endpaper                   
   {\endmode\vfill\supereject}
 
 \def\endit
   {\endpaper\end}


 \def\heading                            
   {\vskip 0.5truein plus 0.1truein      
    \beginparmode \def\\{\par} \parskip=0pt \singlespace \raggedcenter}

 \def\subheading                         
   {\vskip 0.25truein plus 0.1truein     
    \beginlinemode \singlespace \parskip=0pt \def\\{\par}\raggedcenter}

 \def\tag#1$${\eqno(#1)$$}

 \def\align#1$${\eqalign{#1}$$}

 \def\aligntag#1$${\gdef\tag##1\\{&(##1)\cr}\eqalignno{#1\\}$$
   \gdef\tag##1$${\eqno(##1)$$}}

 \def\endaligntag{}

 \def\overset#1\to#2{{\mathop{#2}^{#1}}}
 \def\underset#1\to#2{{\mathop{#2}_{#1}}}

 \def\endreferences{\body}


 \def\ref#1{Ref.~#1}                     
 \def\Ref#1{Ref.~#1}                     
 \def\[#1]{[\cite{#1}]}
 \def\cite#1{{#1}}
 \def\(#1){(\call{#1})}
 \def\call#1{{#1}}
 \def\taghead#1{}
 \def\12{{1\over2}}

 \def\ie{{\it i.e.,\ }}
 
 \def\etal{{\it et al.\ }}

 \def\leaderfill{\leaders\hbox to 1em{\hss.\hss}\hfill}
 \def\twiddle{\lower.9ex\rlap{$\kern-.1em\scriptstyle\sim$}}
 \def\bigtwiddle{\lower1.ex\rlap{$\sim$}}
 \def\gtwid{\mathrel{\raise.3ex\hbox{$>$\kern-.75em\lower1ex\hbox{$\sim$}}}}
 \def\ltwid{\mathrel{\raise.3ex\hbox{$<$\kern-.75em\lower1ex\hbox{$\sim$}}}}
 \def\square{\kern1pt\vbox{\hrule height 1.2pt\hbox{\vrule width 1.2pt\hskip
3pt
    \vbox{\vskip 6pt}\hskip 3pt\vrule width 0.6pt}\hrule height 0.6pt}\kern1pt}
 \def\tdot#1{\mathord{\mathop{#1}\limits^{\kern2pt\ldots}}}
 
 \def\pmb#1{\setbox0=\hbox{#1}%
   \kern-.025em\copy0\kern-\wd0
   \kern  .05em\copy0\kern-\wd0
   \kern-.025em\raise.0433em\box0 }




 \def\references                 
   {\headnn{References}
    \beginparmode                  
    \frenchspacing \parindent=0pt \leftskip=1truecm
    \parskip=8pt plus 3pt \everypar{\hangindent=\parindent}}

\catcode`@=11
\newcount\r@fcount \r@fcount=0
\newcount\r@fcurr
\immediate\newwrite\reffile
\newif\ifr@ffile\r@ffilefalse
\def\w@rnwrite#1{\ifr@ffile\immediate\write\reffile{#1}\fi\message{#1}}

\def\writer@f#1>>{}
\def\referencefile{
\r@ffiletrue\immediate\openout\reffile=\jobname.ref
  \def\writer@f##1>>{\ifr@ffile\immediate\write\reffile%
    {\noexpand\refis{##1} = \csname r@fnum##1\endcsname = %
     \expandafter\expandafter\expandafter\strip@t\expandafter%
     \meaning\csname r@ftext\csname r@fnum##1\endcsname\endcsname}\fi}%
  \def\strip@t##1>>{}}

\def\citeall#1{\xdef#1##1{#1{\noexpand\cite{##1}}}}
\def\cite#1{\each@rg\citer@nge{#1}} 

\def\each@rg#1#2{{\let\thecsname=#1\expandafter\first@rg#2,\end,}}
\def\first@rg#1,{\thecsname{#1}\apply@rg} 
\def\apply@rg#1,{\ifx\end#1\let\next=\relax
\else,\thecsname{#1}\let\next=\apply@rg\fi\next}

\def\citer@nge#1{\citedor@nge#1-\end-} 
\def\citer@ngeat#1\end-{#1}
\def\citedor@nge#1-#2-{\ifx\end#2\r@featspace#1 
  \else\citel@@p{#1}{#2}\citer@ngeat\fi} 
\def\citel@@p#1#2{\ifnum#1>#2{\errmessage{Reference range #1-#2\space is bad.}%
    \errhelp{If you cite a series of references by the notation M-N, then M and
    N must be integers, and N must be greater than or equal to M.}}\else%
 {\count0=#1\count1=#2\advance\count1
by1\relax\expandafter\r@fcite\the\count0,%
  \loop\advance\count0 by1\relax
    \ifnum\count0<\count1,\expandafter\r@fcite\the\count0,%
  \repeat}\fi}

\def\r@featspace#1#2 {\r@fcite#1#2,} 
\def\r@fcite#1,{\ifuncit@d{#1}
    \newr@f{#1}%
    \expandafter\gdef\csname r@ftext\number\r@fcount\endcsname%
                     {\message{Reference #1 to be supplied.}%
                      \writer@f#1>>#1 to be supplied.\par}%
 \fi%
 \csname r@fnum#1\endcsname}
\def\ifuncit@d#1{\expandafter\ifx\csname r@fnum#1\endcsname\relax}%
\def\newr@f#1{\global\advance\r@fcount by1%
    \expandafter\xdef\csname r@fnum#1\endcsname{\number\r@fcount}}

\let\r@fis=\refis   
\def\refis#1#2#3\par{\ifuncit@d{#1}
   \newr@f{#1}%
   \w@rnwrite{Reference #1=\number\r@fcount\space is not cited up to now.}\fi%
  \expandafter\gdef\csname r@ftext\csname r@fnum#1\endcsname\endcsname%
  {\writer@f#1>>#2#3\par}}

\def\ignoreuncited{
   \def\refis##1##2##3\par{\ifuncit@d{##1}%
     \else\expandafter\gdef\csname r@ftext\csname
r@fnum##1\endcsname\endcsname%
     {\writer@f##1>>##2##3\par}\fi}}

\def\r@ferr{\endreferences\errmessage{I was expecting to see
\noexpand\endreferences before now;  I have inserted it here.}}
\let\r@ferences=\references
\def\references{\r@ferences\def\endmode{\r@ferr\par\endgroup}}

\let\endr@ferences=\endreferences
\def\endreferences{\r@fcurr=0
  {\loop\ifnum\r@fcurr<\r@fcount
    \advance\r@fcurr by 1\relax\expandafter\r@fis\expandafter{\number\r@fcurr}%
    \csname r@ftext\number\r@fcurr\endcsname%
  \repeat}\gdef\r@ferr{}\endr@ferences}


\let\r@fend=\endpaper\gdef\endpaper{\ifr@ffile
\immediate\write16{Cross References written on []\jobname.REF.}\fi\r@fend}

\catcode`@=12

\citeall\refto  
\citeall\ref  %
\citeall\Ref  %

\ignoreuncited

\catcode`@=11
\newcount\tagnumber\tagnumber=0

\immediate\newwrite\eqnfile
\newif\if@qnfile\@qnfilefalse
\def\write@qn#1{}
\def\writenew@qn#1{}
\def\w@rnwrite#1{\write@qn{#1}\message{#1}}
\def\@rrwrite#1{\write@qn{#1}\errmessage{#1}}

\def\taghead#1{\gdef\t@ghead{#1}\global\tagnumber=0}
\def\t@ghead{}

\expandafter\def\csname @qnnum-3\endcsname
  {{\t@ghead\advance\tagnumber by -3\relax\number\tagnumber}}
\expandafter\def\csname @qnnum-2\endcsname
  {{\t@ghead\advance\tagnumber by -2\relax\number\tagnumber}}
\expandafter\def\csname @qnnum-1\endcsname
  {{\t@ghead\advance\tagnumber by -1\relax\number\tagnumber}}
\expandafter\def\csname @qnnum0\endcsname
  {\t@ghead\number\tagnumber}
\expandafter\def\csname @qnnum+1\endcsname
  {{\t@ghead\advance\tagnumber by 1\relax\number\tagnumber}}
\expandafter\def\csname @qnnum+2\endcsname
  {{\t@ghead\advance\tagnumber by 2\relax\number\tagnumber}}
\expandafter\def\csname @qnnum+3\endcsname
  {{\t@ghead\advance\tagnumber by 3\relax\number\tagnumber}}

\def\equationfile{%
  \@qnfiletrue\immediate\openout\eqnfile=\jobname.eqn%
  \def\write@qn##1{\if@qnfile\immediate\write\eqnfile{##1}\fi}
  \def\writenew@qn##1{\if@qnfile\immediate\write\eqnfile
    {\noexpand\tag{##1} = (\t@ghead\number\tagnumber)}\fi}
}

\def\callall#1{\xdef#1##1{#1{\noexpand\call{##1}}}}
\def\call#1{\each@rg\callr@nge{#1}}

\def\each@rg#1#2{{\let\thecsname=#1\expandafter\first@rg#2,\end,}}
\def\first@rg#1,{\thecsname{#1}\apply@rg}
\def\apply@rg#1,{\ifx\end#1\let\next=\relax%
\else,\thecsname{#1}\let\next=\apply@rg\fi\next}

\def\callr@nge#1{\calldor@nge#1-\end-}
\def\callr@ngeat#1\end-{#1}
\def\calldor@nge#1-#2-{\ifx\end#2\@qneatspace#1 %
  \else\calll@@p{#1}{#2}\callr@ngeat\fi}
\def\calll@@p#1#2{\ifnum#1>#2{\@rrwrite{Equation range #1-#2\space is bad.}
\errhelp{If you call a series of equations by the notation M-N, then M and
N must be integers, and N must be greater than or equal to M.}}\else%
 {\count0=#1\count1=#2\advance\count1
by1\relax\expandafter\@qncall\the\count0,%
  \loop\advance\count0 by1\relax%
    \ifnum\count0<\count1,\expandafter\@qncall\the\count0,%
  \repeat}\fi}

\def\@qneatspace#1#2 {\@qncall#1#2,}
\def\@qncall#1,{\ifunc@lled{#1}{\def\next{#1}\ifx\next\empty\else
  \w@rnwrite{Equation number \noexpand\(>>#1<<) has not been defined yet.}
  >>#1<<\fi}\else\csname @qnnum#1\endcsname\fi}

\let\eqnono=\eqno
\def\eqno(#1){\tag#1}
\def\tag#1$${\eqnono(\displayt@g#1 )$$}

\def\aligntag#1\endaligntag
  $${\gdef\tag##1\\{&(##1 )\cr}\eqalignno{#1\\}$$
  \gdef\tag##1$${\eqnono(\displayt@g##1 )$$}}

\def\eqalignno#1{\displ@y \tabskip\centering
  \halign to\displaywidth{\hfil$\displaystyle{##}$\tabskip\z@skip
    &$\displaystyle{{}##}$\hfil\tabskip\centering
    &\llap{$\displayt@gpar##$}\tabskip\z@skip\crcr
    #1\crcr}}

\def\displayt@gpar(#1){(\displayt@g#1 )}

\def\displayt@g#1 {\rm\ifunc@lled{#1}\global\advance\tagnumber by1
        {\def\next{#1}\ifx\next\empty\else\expandafter
        \xdef\csname @qnnum#1\endcsname{\t@ghead\number\tagnumber}\fi}%
  \writenew@qn{#1}\t@ghead\number\tagnumber\else
        {\edef\next{\t@ghead\number\tagnumber}%
        \expandafter\ifx\csname @qnnum#1\endcsname\next\else
        \w@rnwrite{Equation \noexpand\tag{#1} is a duplicate number.}\fi}%
  \csname @qnnum#1\endcsname\fi}

\def\ifunc@lled#1{\expandafter\ifx\csname @qnnum#1\endcsname\relax}

\let\@qnend=\end\gdef\end{\if@qnfile
\immediate\write16{Equation numbers written on []\jobname.EQN.}\fi\@qnend}

\catcode`@=12

\catcode`@=11 
\def\lsim{\mathrel{\mathpalette\@versim<}}
\def\gsim{\mathrel{\mathpalette\@versim>}}
\def\@versim#1#2{\lower0.2ex\vbox{\baselineskip\z@skip\lineskip\z@skip
  \lineskiplimit\z@\ialign{$\m@th#1\hfil##\hfil$\crcr#2\crcr\sim\crcr}}}
\catcode`@=12 

\def\E#1 {\times 10^{#1}}
\def\pagenumbers{\pageno=1\global\footline={\hfil\twelverm-- \folio\ --\hfil}}

\newbox\charbox
\newbox\slabox
\def\sla#1{{\setbox\charbox=\hbox{$#1$}
           \setbox\slabox=\hbox{$/$}
           \dimen\charbox=\ht\slabox
           \advance\dimen\charbox by -\dp\slabox
           \advance\dimen\charbox by -\ht\charbox
           \advance\dimen\charbox by \dp\charbox
           \divide\dimen\charbox by 2
           \raise-\dimen\charbox\hbox to \wd\charbox{\hss/\hss}
           \llap{$#1$}
          }}

\catcode`@=11 

\newcount\chapterno
\global\chapterno=0
\newcount\appendixno
\global\appendixno=0
\newcount\headno
\global\headno=0
\newcount\subheadno
\global\subheadno=0
\newcount\figno
\global\figno=0
\newcount\tabno
\global\tabno=0
\newcount\noheadline
\newcount\curr@head
\newcount\c@ntentsno
\global\c@ntentsno=0
\newcount\str@pno
\newcount\c@ntents
\global\c@ntents=0
\newcount\currc@nt

\def\bighead#1{\vfill\supereject
\global\advance\c@ntentsno by1
\xdef\caname{#1}
\expandafter\xdef\csname h@name\number\c@ntentsno\endcsname
{B\noexpand\else #1}
\immediate\write16{\number\c@ntentsno :\caname}
{\vglue 3em \myraggedcenter\bigfont #1 \vglue 2em\hrule\vglue 2em}
\noheadline=\pageno
\mark{\number\c@ntentsno}
}

\def\chapter#1{\vfill\supereject
\iftwoside\ifodd\pageno\else\noheadline=\pageno\null\vfill\eject\fi\fi
\global\advance\chapterno by1
\global\advance\c@ntentsno by1
\global\headno=0
\global\subheadno=0
\global\figno=0
\global\tabno=0
\taghead{\number\chapterno .}
\xdef\caname{Chapter \number\chapterno : #1}
\xdef\cano{\number\chapterno}
\expandafter\xdef\csname h@name\number\c@ntentsno\endcsname
{C\noexpand\else Chapter \number\chapterno : #1}
\immediate\write16{\number\c@ntentsno :\caname}
{\vglue 3em
\myraggedcenter\bigfont Chapter \number\chapterno
\bigskip\hrule\bigskip #1 \vglue 2em}
\noheadline=\pageno
\mark{\number\c@ntentsno}
}

\def\nolet#1{\ifcase#1{This shouldn't happen}\or A\or B\or C\or D%
\or E\or F\or G\or H\or I\or J\or K\or L\or M\or N\or O\or P\or Q%
\or R\or S\or T\or U\or V\or W\or X\or Y\or Z\else{\Omega}\fi}

\def\appendix#1{\vfill\supereject
\iftwoside\ifodd\pageno\else\noheadline=\pageno\null\vfill\eject\fi\fi
\global\advance\appendixno by1
\global\advance\c@ntentsno by1
\global\headno=0
\global\subheadno=0
\global\figno=0
\global\tabno=0
\taghead{\nolet\appendixno .}
\xdef\caname{Appendix \nolet\appendixno : #1}
\xdef\cano{\nolet\appendixno}
\expandafter\xdef\csname h@name\number\c@ntentsno\endcsname
{A\noexpand\else Appendix \nolet\appendixno : #1}
\immediate\write16{\number\c@ntentsno :\caname}
{\vglue 3em
\myraggedcenter\bigfont Appendix \nolet\appendixno
\bigskip\hrule\bigskip #1 \vglue 2em}
\noheadline=\pageno
\mark{\number\c@ntentsno}
}

\gdef\caname{}
\gdef\cano{}

\headline={\hfil}

\footline={\hfil\twelverm-- \folio\ --\hfil}

\gdef\s@s{ }

\output={
 \shipout\vbox{\columnbox}
 \advancepageno
\ifnum\outputpenalty>-20000 \else\dosupereject\fi}

\def\columnbox{\vbox{
\ifnum\firstmark=0\else
\ifnum\topmark=\botmark
\else\curr@head=\firstmark
\loop
\ifnum\curr@head>\c@ntents
\else
\ifnum\pageno=\csname c@ntentpage\number\curr@head\endcsname\else
\fi\fi
\edef\contype{\iftrue\csname h@name\number\curr@head\endcsname\fi}
\ifnum\curr@head <\botmark \advance\curr@head by 1
\repeat
\fi\fi
\leftline{\vbox{\makeheadline\pagebody\makefootline}}}}

\def\contentspage#1#2#3#4{
\global\advance\c@ntents by1
\ifnum #1=\c@ntents
\expandafter\gdef\csname c@ntenttype#1\endcsname{#2}
\expandafter\gdef\csname c@ntenttext#1\endcsname{#3}
\expandafter\gdef\csname c@ntentpage#1\endcsname{#4}
\else
\message{Your contents page is garbled (no.#1) ... for a happy life correct
it}
\global\advance\c@ntents by-1
\fi
}

\def\bdotfil{\leaders\hbox to 1.5em{\hss.\hss}\hfil}

\def\contents{\ifnum\c@ntents=0
\message{Eh up .... where's the Contents page got to ?}
\else
\bighead{Contents}
{\currc@nt=1\loop
\message{\number\currc@nt}
\edef\c@ntype{\csname c@ntenttype\number\currc@nt\endcsname}
\if\c@ntype A
 \vskip 1cm\goodbreak\noindent
 {\bf\csname c@ntenttext\number\currc@nt\endcsname}
 \hfill\csname c@ntentpage\number\currc@nt\endcsname\fi
\if\c@ntype B
 \vskip 1cm\goodbreak\noindent
 {\bf\csname c@ntenttext\number\currc@nt\endcsname}
 \bdotfil\csname c@ntentpage\number\currc@nt\endcsname\fi
\if\c@ntype C
 \vskip 1cm\goodbreak\noindent
 {\bf\csname c@ntenttext\number\currc@nt\endcsname}
 \hfill\csname c@ntentpage\number\currc@nt\endcsname\fi
\if\c@ntype H
 \noindent\hbox{\hskip 1em}
 \csname c@ntenttext\number\currc@nt\endcsname
 \bdotfil\csname c@ntentpage\number\currc@nt\endcsname\fi
\if\c@ntype S
 \noindent\hbox{\hskip 2em}
 \csname c@ntenttext\number\currc@nt\endcsname
 \bdotfil\csname c@ntentpage\number\currc@nt\endcsname\fi
\break
\ifnum\currc@nt <\c@ntents \advance\currc@nt by 1
\repeat}\fi
}

\mark{0}

\def\head#1{                    
  \global\advance\headno by1
  \global\advance\c@ntentsno by1
  \global\subheadno=0
  \vskip 0.5truein\goodbreak    
\expandafter\xdef\csname h@name\number\c@ntentsno\endcsname
   {H\noexpand\else \number\headno\ #1}
   \immediate\write16{\number\c@ntentsno :#1}
    \taghead{\number\headno .}
\goodbreak
\vbox{
    \leftline{\bf\number\headno\ #1}
    \nobreak\vskip 0.25truein\nobreak
     }
    \mark{\number\c@ntentsno}
}

\def\headnn#1{
\vskip 0.5truein\goodbreak
\vbox{
    \leftline{\bf #1}
    \nobreak\vskip 0.25truein\nobreak
     }
}
\def\subhead#1{                    
  \global\advance\subheadno by1
  \global\advance\c@ntentsno by1
   \vskip 0.25truein\goodbreak    
\expandafter\xdef\csname h@name\number\c@ntentsno\endcsname
   {S\noexpand\else \cano .\number\headno.\number\subheadno\ #1}
   \immediate\write16{\number\c@ntentsno :#1}
\goodbreak
\vbox{
    \leftline{\bf\cano .\number\headno .\number\subheadno \ #1}
    \nobreak\vskip 0.1truein\nobreak
     }
    \mark{\number\c@ntentsno}
}

\def\fig#1#2{
	\global\advance\figno by 1
	\expandafter\xdef\csname f@g#1\endcsname{\number\figno}
	{\par\myraggedcenter Fig. \number\figno{} #2\smallskip}
}

\def\rfig#1{\hbox{Fig. \csname f@g#1\endcsname}}

\def\figref#1#2{\expandafter\xdef\csname f@g#1\endcsname{#2}}

\def\table#1#2{
	\global\advance\tabno by 1
	\expandafter\xdef\csname t@b#1\endcsname{\number\figno}
	{\myraggedcenter{\it Table} \number\tabno{} #2\smallskip}
}

\def\rtable#1{\hbox{{\it Table} \csname t@b#1\endcsname}}

\def\tablref#1#2{\expandafter\xdef\csname t@b#1\endcsname{#2}}

\def\first#1#2{{\str@pno=#1\expandafter\str@p#2 \end{} }}
\def\str@p#1 {\ifx\end#1\let\next=\relax%
\else\advance\str@pno by -1
\ifnum\str@pno=-1$...$\def\next##1\end{}%
\else #1 \let\next=\str@p\fi\fi \next}

\catcode`@=12 

 \def\myraggedcenter{\leftskip=0em plus 12em \rightskip=\leftskip
   \parindent=0pt \parfillskip=0pt \spaceskip=.3333em \xspaceskip=.5em
   \pretolerance=9999 \tolerance=9999
   \hyphenpenalty=9999 \exhyphenpenalty=9999 }

%
%
\def\frac#1#2{{\scriptstyle#1\over\scriptstyle#2}}
\def\half{{\frac12}}

\def\fourth{{\frac14}}

\def\Tr#1{\hbox{Tr}\noexpand\left[#1\noexpand\right]}

\def\ddot#1#2{\thinspace #1\!\cdot\! #2\thinspace}

\def\LEPI/{\hbox{\rm LEP I}}
\def\LEPII/{\hbox{\rm LEP I$\!$I}}
\def\GeV{{\rm\, GeV}}

\font\circle=lcircle10
\def\trcorner{\hbox{\raise 2.9pt\hbox{\circle\char10}\kern -10pt}}
\def\barrow{\trcorner\to}

\def\gammav{\gamma{\!\raise -0.5em\hbox{$\scriptstyle 5$}}}

\def\star{^{\scriptscriptfont0=\fiverm \scriptscriptstyle ( \scriptstyle *
\scriptscriptstyle ) }}
\def\BW{\hbox{BW}}

\nopagenumbers
\hbox{\font\fortssbx=cmssbx10 scaled \magstep2
\epsfysize=0.45in
\epsfbox{uwlogo.eps}
\hskip 2cm\raise 12pt\hbox{\fortssbx University of Wisconsin -- Madison}
}

\hfill\vbox{\rm\hbox{MAD/PH/861}
\hbox{June 1995}}
\title
Finite width effects in Higgs decays as a means of measuring massive %
particle widths
\bigskip
\author
D.J. Summers%
\footnote*{Email address: summers @ phenxr.physics.wisc.edu}
\affil
Department of Physics,
University of Wisconsin -- Madison,
1150 University Avenue,
Madison,
WI 53706,
U.S.A.
\bigskip

\abstract

We calculate decays of a Standard Model Higgs boson to a virtual
massive particle and discuss how this depends on the massive particle
total width. If the partial width of Higgs to a virtual massive
particle can be measured this gives a measurement of that massive
particle's width. We discuss how one would go about measuring these
partial widths of a Higgs experimentally, and how this could lead to a
measurement of the $W$ boson and $t$ quark width. For the latter extreme
dependence on the Higgs mass and the small $H\to tt^*$ branching ratios
mean that little can be learnt about the $t$ quark width. For the
former there is also large dependence on the Higgs mass; however this
can be removed by taking the ratio of $H\to WW^*$ decays to
$H\to ZZ^*$ decays. This ratio also has the advantage of being fairly
insensitive to physics beyond the Standard Model. Unfortunately, for
Higgs masses of interest the $H\to ZZ^*$ branching ratio is small
enough that we require many 1000's of tagged Higgs decays before an
accurate measurement of the $W$ width can be made. This is likely to
be hard experimentally.

\bigskip
\noindent {\bf PACS:} 14.80.Bn 13.38.-b

\endpage
\body

\pagenumbers
\head{Introduction}

The $W$ and $Z$ bosons that we observe in high energy particle
colliders are massive objects. This mass does not enter our
theoretical models at a fundamental level, but instead is typically
generated by some spin zero object. The physical existence of this
spin zero object is among the most important questions that we
have about physical reality -- and if this particle is discovered
then its properties will become of prime interest. In the Standard
Model this spin zero object is also a fundamental field of the
theory, the Higgs boson.
In this paper we examine how the properties of this Higgs boson may be
used to probe other aspects of Standard Model physics; in particular
we discuss how Higgs decays to off-shell massive particles may be used
as a probe of that particles width.

If a particle has a mass below the threshold to decay to on-shell
massive particles, then that decay can still occur through the decay
to off-mass-shell particles; although the decay rate is usually very
small due to the off-mass shell propagator. However
in the Standard Model the Higgs boson is responsible for generating
all particle masses. So the more massive the particle the
stronger the Higgs couples to that particle. This means that Higgs
decays to off-shell massive particles, although suppressed by the
off-shell propagator, are enhanced by the strong Higgs coupling. Thus
Higgs decays to off-shell particles can be appreciable, this can
be seen especially in the case $H\to W\star W\star$ where the
branching ratio is above 10\% for Higgs masses as low as $115\GeV$
despite this forcing a $W$ at least $45\GeV$ off mass shell.

It is interesting to consider how a two stage decay takes place; that
is a decay that proceeds via an intermediate particle,
$$
A \to B \to C \qquad .\tag
$$
At leading order (LO) the decay rate for this looks like
$$
\Gamma(A\to B\to C) \sim
\int {d(p^2) \over (p^2-m_B^2)^2 + m_B^2 \Gamma_B^2}
\int d({\rm LIPS}_{B\to C}) |{\cal M}|^2 \qquad .\tag
$$
Now when the
$B$ width, $\Gamma_B$, is small the integral over $p^2$ can be done.
When we are above threshold the narrow width approximation gives
$$
\Gamma(A) \sim {\pi \over m_B \Gamma_B}
\int d({\rm LIPS}_{B\to C}) |{\cal M}|^2 \qquad .\tag
$$
This appears to diverge in the limit $\Gamma_B \to 0$; however the
integral over the $B$ to $C$ phase space gives a term
$$
\int d({\rm LIPS}_{B\to C}) |{\cal M}|^2 \sim \Gamma_B^{\rm LO}
{\rm Br}(B\to C) \qquad ,\tag
$$
and so in the narrow width approximation we find
$$
\Gamma(A\to B\to C) \simeq \Gamma(A \to B)
{\Gamma_B^{\rm LO} \over \Gamma_B} {\rm Br}(B\to C) \qquad .\tag
$$
This is the result we expect as long as
$\Gamma_B^{\rm LO} = \Gamma_B$, that is that the $B$ width in the
Breit-Wigner propagator is the same as the width that comes from
integrating the matrix element over the $B$ decay phase space. We could
of course choose $\Gamma_B$ equal to $\Gamma_B^{\rm LO}$; however as
we include higher order corrections to the decay of $B$ we expect that
the $\Gamma_B^{\rm LO}$ in the numerator will tend towards the
physical value of the width, $\Gamma_B$. As such in this work we will
always replace the width in the numerator by the physical width,
$\Gamma_B$.

If we now consider the case where the decay of $A$ via an on-shell $B$
is kinematically forbidden, then $A$ can still decay via a virtual
$B^*$,
$$
A \to B^* \to C \qquad .\tag
$$
Then in the narrow width approximation the integral over the Breit
Wigner becomes
$$
\int {d(p^2) \over (p^2-m_B^2)^2 + m_B^2 \Gamma_B^2}
\to \int {d(p^2) \over (p^2-m_B^2)^2 } \qquad ,\tag
$$
so this integral is no longer proportional to $1/\Gamma_B$;
however we still get a $\Gamma_B$ in the numerator of the width from
the integral of the matrix element over the $B$ phase space. This
means that
$$
\Gamma(A \to B^*) \sim \Gamma(B) \qquad , \tag
$$
and so measuring the decay width of $A$ into a virtual $B$ gives us
information about the $B$ width, or more strictly the running $B^*$
width, which is related to the $B$ width.

In this paper we calculate the decays of the Standard Model Higgs
boson that proceed via a massive virtual particle. In the Standard
Model there are 3 very massive
particles, the $W$ and $Z$ bosons and the $t$ quark; we consider
all Higgs decays that involve these massive particles,
$$
\eqalignno{
H &\to Z\star \gamma \to f_1 \bar f_1 \gamma &(Zg) \cr
H &\to Z\star Z\star \to f_1 \bar f_1 \; f_2 \bar f_2 &(ZZ) \cr
H &\to W\star W\star \to f_1 \bar f_2 \; f_3 \bar f_4 &(WW) \cr
H &\to t\star \bar t\star \to b W\star \bar b W\star
   \to b f_1 \bar f_2 \; \bar b f_3 \bar f_4 &(tt) \cr
}
$$
The rates for processes \(ZZ,WW,tt) have been calculated before in
\ref{RJNP}, the rate for process \(Zg) has not previously been
published\refto{thesis}.

\head{Breit--Wigner propagators}

As we are interested in measuring massive particle widths in this
paper, and a large dependence on the particles width arises from the
Breit-Wigner propagator, we need to consider the form of the
Breit-Wigner propagator that we use. If we start with a bare massive
spin 0 particle propagator and we sum an arbitrary number of one particle
irreducible insertions we have,
$$
\eqalign{
{\rm Prop} &=   {1 \over p^2 - m_0^2}
             + {1 \over p^2 - m_0^2} \Pi(p^2) {1 \over p^2 - m_0^2}
             + {1 \over p^2 - m_0^2} \Pi(p^2) {1 \over p^2 - m_0^2}
                        \Pi(p^2) {1 \over p^2 - m_0^2}
             + \ldots \cr
           &= {1 \over p^2 - m_0^2 - \Pi(p^2)} \cr
           &= {1 \over p^2 - (m_0^2 + {\rm Re}(\Pi(p^2)))
                               - i {\rm Im}(\Pi(p^2))} \cr
           &= {1 \over p^2 - m_R^2(p^2) - i {\rm Im}(\Pi(p^2))} \qquad ,\cr
} \tag
$$
where the real part of the one particle irreducible diagrams have been
reabsorbed into the definition of the particle mass to give a running
mass. The running mass is then related to the physical pole mass
through the relationship
$$
m_R^2(m^2)=m^2 \qquad .\tag
$$
In this work we will not calculate the real part of the one particle
irreducible diagrams at all, and will always use a fixed mass,
$$
m_R^2(p^2) = m^2 \qquad .\tag
$$
Now the imaginary part of the one particle irreducible diagrams is
related to the total width through the optical theorem
$$
\eqalign{
{\rm Im}(\Pi(p^2)) &= -{1\over 2}
               \int d({\rm LIPS}) |{\cal M}_{\rm decay}|^2 \cr
	&= -m \Gamma(p^2)\qquad ,\cr
}\tag
$$
where we have defined the running width as
$$
\Gamma(q^2) = {1 \over 2 m}
     \int d({\rm LIPS}_{\rm q^2}) |{\cal M}_{\rm decay}|^2
\qquad .\tag
$$
Notice that we use $1 / 2m$ as the flux factor for all $q^2$ values,
whereas $\int d({\rm LIPS}_{\rm q^2}) |{\cal M}_{\rm decay}|^2$ is
evaluated for a particle with $p^2 = q^2$ to decay.

This gives the Breit--Wigner propagator
$$
{\rm Prop}_0 = {1 \over p^2 - m^2 + i m \Gamma(p^2)}
\qquad .\tag
$$
If we consider the propagator of a massive spin 1 gauge boson, then
if that particle only decays to massless fermions (as we will
consider in this paper) then we only have a contribution from the
transverse part of the propagator. This gives the form of the
propagator as
$$
{\rm Prop}_1 = {-g^{\mu\nu} + p^\nu p^\nu / p^2
          \over p^2 - m^2 + i m \Gamma(p^2)}
\qquad ,\tag
$$
where the $p^\nu p^\nu / p^2 $ term in the numerator always cancels on
massless fermions. If the spin 1 particle only decays into massless
particles then, neglecting the running of coupling constants,
we know that $m \Gamma(p^2) \sim p^2$ and so we can write the
propagator in terms of the on-mass-shell width,
$$
{\rm Prop}_1 = {-g^{\mu\nu} + p^\nu p^\nu / p^2
          \over p^2 - m^2 + i p^2 \Gamma(m^2)/m}
\qquad .\tag
$$

The Breit Wigner for a massive fermion is worth considering in more
detail. The imaginary part of the one particle irreducible diagrams
come from the diagrams where the heavy quark decays into a light quark
through the emission of a $W$ boson.
The one particle irreducible insertion, without the heavy quark
propagators,  is given by
$$
\Pi = B(p^2) \sla p (1-\gammav) \qquad ,\tag
$$
where $p$ is the heavy quark momentum and B is a scalar function of $p^2$
only. Multiplying by $(\sla p + m)$ and
taking the trace gives
$$
4 p^2 {\rm Im}(B)= {\rm Im}(B) {\rm Tr}((\sla p + m)\sla p (1-\gammav))
        = {\rm Im}({\rm Tr}(( \sla p + m) \Pi))
	= - 2 m \Gamma
\qquad ,\tag
$$
where the last equality is given by the optical theorem. This gives
$$
{\rm Im}(B) = - {m \Gamma \over 2 p^2} \qquad .\tag
$$
After resumming the one particle irreducible diagrams the heavy quark
propagator is given by
$$
\eqalign{
{\rm Prop}_{1/2} &= {1 \over \sla p - m}
	+ {1 \over \sla p - m} \Pi {1 \over \sla p - m}
	+ {1 \over \sla p - m} \Pi {1 \over \sla p - m}
		\Pi {1 \over \sla p - m}
	+ \ldots \cr
	&= {1 \over \sla p - m - B \sla p (1-\gammav)} \cr
	&= {\sla p + m - B \sla p (1-\gammav) \over p^2(1-2B) -m^2}
\qquad .\cr
}\tag
$$
Now if either end of the heavy quark propagator couples onto a $W$
boson then the $\sla p (1-\gammav)$ in the numerator cancels
directly. As in this paper we decay all top quarks, and so one end of
the propagator is always
coupled to a $W$, we drop the $\sla p (1-\gammav)$ term in the
numerator. Then
the the imaginary part of $B$ gives the fermion Breit--Wigner as
$$
{\rm Prop}_{1/2} =
	{\sla p + m \over p^2  -m^2 + i m \Gamma(p^2) }
\qquad .\tag
$$
Although this is the naive form of the Breit--Wigner fermion
propagator we notice that it disagrees with the form of the
Breit--Wigner fermion propagator given in \ref{fermprop}.
Also notice that this form of the
propagator is different from \ref{RJNP} where they use $1/2\sqrt{q^2}$
as the flux factor in the definition of an off-shell decay width,
whereas we use $1/2m$.

\head{Higgs decay widths}
\topinsert
\centerline{
 \epsfxsize=6.5in\epsfbox[40 115 675 205]{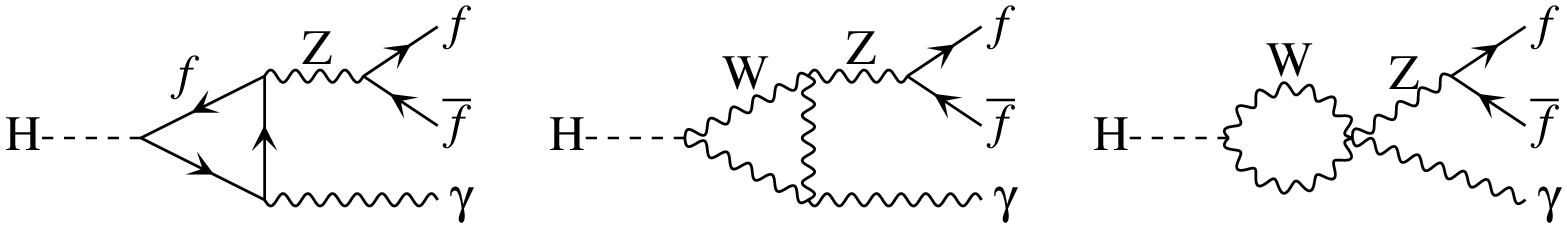}
}
\fig{HZgamma}{The Feynman diagrams for the process
$H \to Z^* \gamma \to f \bar f \gamma$.}
\endinsert
In this section we give the decay rates for a Higgs to decay in the
channels (\call{Zg}--\call{tt}). 
Now the decay \(Zg) does not occur at tree level, and so the lowest
order diagrams are at the 1 loop level and shown in \rfig{HZgamma}.
At the same order in perturbation theory there are other Feynman
diagrams that also contribute to the decay $H \to f \bar f \gamma$,
that do not proceed via a single $Z$ boson. In this work we will
calculate just the diagrams associated with the decay
$H\to Z^*\gamma$, even though this does not give the full rate for
$H\to f \bar f \gamma$, or is even gauge invariant with respect to the
$SU(2)$ gauge. This means that if we find any interesting physics
associated with the decay $H\to Z^* \gamma$ then a more complete
calculation needs to be done.
As the result for \(Zg) has not been
previously published we give a fairly complete derivation of this
result. We also give the partial widths for (\call{ZZ}--\call{tt}),
although
these have been calculated before \refto{RJNP}, we give the results again
here. Our results differ slightly from \ref{RJNP}, as we have defined
our running widths in a slightly different way, and hence use
a different form of the running width in the Breit--Wigner propagator.
Our form of the Breit--Wigner propagator is motivated by the use of
the optical theorem and so we expect it to be more accurate than that
used in \ref{RJNP}.

The Feynman diagrams for
$H \to Z\star \gamma \to f_1 \bar f_1 \gamma $
are shown in \rfig{HZgamma}, it is
convenient with this process to split the calculation up into two
halves, into the process $H \to Z^* \gamma$ followed by the process
$Z^* \to f \bar f$. The calculation of the process $H \to Z^* \gamma$
is identical to the calculation of $H \to Z\gamma$, which has been
done many times before. See for example \ref{HHG}. Here we do not
repeat the calculation but just quote the results. The effective
coupling for $HZ\gamma$ vertex is given by
$$
{\cal M}_{HZ\gamma}^{\mu\nu}=
A (p_Z^\mu p_\gamma^\mu - \ddot{p_Z}{p_\gamma} g^{\mu\nu}) \qquad ,\tag
$$
where $A$ has the form \refto{HHG}
$$
A = {\alpha g \over 4 \pi M_W}(A_F + A_W) \tag
$$
and $A_F$ and $A_W$ are given by
$$
A_F = \sum_{\rm fermions} n_c { - 2 e_f (T_f^3 - 2e_f \sin^2\theta_W )
                                \over \sin\theta_W\cos\theta_W }
      [I_1(\tau_f,\lambda_f) - I_2(\tau_f,\lambda_f)] \tag
$$
and
$$
\eqalign{
A_W=& -\cot\theta_W \Biggl\{ 4 ( 3-\tan^2 \theta_W )
                                 I_2(\tau_W,\lambda_W) \cr
     &+ \left[ \left( 1 + {2\over \tau_W} \right) \tan^2\theta_W
             - \left(5 + {2 \over \tau_W} \right) \right]
                    I_1(\tau_W,\lambda_W) \Biggr\} \qquad ,\cr}\tag
$$
where,
$$
\tau_f \equiv {4m_f^2 \over m_H^2} \qquad
\lambda_f \equiv {4m_f^2 \over p_Z^2} \qquad
\tau_W \equiv {4m_W^2 \over m_H^2} \qquad
\lambda_W \equiv {4m_W^2 \over p_Z^2} \qquad ,\tag
$$
with,
$$
\eqalign{
I_1(a,b)&={ab \over 2 (a-b)} + {a^2b^2 \over 2 (a-b)^2}[f(a)-f(b)]
         + {a^2 b \over (a-b)^2}[g(a)-g(b)] \cr
I_2(a,b)&= - {ab \over 2(a-b)}[f(a)-f(b)] \cr}
$$$$
f(\tau)= \cases{\left[\sin^{-1}\left(\sqrt{1/\tau}\right)\right]^2
                                    &if $\tau\ge 1$,\cr
                 -\fourth [\ln(\eta_+/\eta_-) - i \pi ]^2
                                    &if $\tau< 1$,} \tag
$$$$
g(\tau)=\cases{ \sqrt{\tau-1} \sin^{-1} (1/\sqrt\tau)&if $\tau\ge 1$,\cr
                \half\sqrt{\tau-1} [\ln(\eta_+/\eta_-) - i \pi ]
                                   &if $\tau< 1$,} \tag
$$$$
\eta_\pm \equiv ( 1 \pm \sqrt{ 1 -\tau }) \qquad .\tag
$$
This leads to the matrix element squared
$$
|{\cal M}|^2 = 8 (g_V^2 + g_A^2) {|A|^2 g_Z^2 \over \BW_Z(p_Z^2) }
               \ddot{p_f}{p_{\bar f}}
  \left((\ddot{p_\gamma}{p_f})^2 +(\ddot{p_\gamma}{p_{\bar f}})^2\right)
\qquad ,\tag
$$
where
$$
\BW_V(p_V^2) \equiv (p_V^2 -M_V^2)^2 +M_V^2 \Gamma_V^2(p_V^2) \qquad .\tag
$$
The partial width is given by
$$
d\Gamma={(2\pi)^4} {1 \over 2 M_H}
        \delta^4\left(p_H-\sum p_{\rm final}\right)
        |{\cal M}|^2 \prod {d^3 p_{\rm final}
        \over 2 E_{\rm final} (2\pi)^3} \qquad .\tag
$$
Most of these integrations can be done analytically, leaving
the integration over the $Z$ line shape
$$
\Gamma(H \to Z^* \gamma)
= {\Gamma_Z M_H^3 \over 32 \pi^2 M_Z} \int_0^{M_H^2} dp_Z^2
  \left( 1 -{p_Z^2 \over M_H^2}\right)^3
  {A^2 p_Z^2 \over \BW_Z(p_Z^2)}
\qquad ,\tag
$$
where we have replaced the $M_Z /( 12\pi) g_Z^2(g_V^2+g_A^2)$ in the
numerator with the on-shell non-running $Z$ width. This last integral
over the $Z$ line shape is best done numerically.

The partial width for $H \to V\star V\star$ is given by
$$
\eqalign{
\Gamma(H \to V^* V^*) =
{ \Gamma_V^2 \over M_V^2}  {g_{HVV}^2   \over   64 \pi^3 M_H }
 &\int {d(p_{V_1}^2) \over  \BW_{V_1}(p_{V_1}^2)}
      {d(p_{V_2}^2) \over  \BW_{V_2}(p_{V_2}^2)}
((M_H^2-p_{V_1}^2-p_{V_2}^2)^2 + 8 p_{V_1}^2 p_{V_2}^2) \cr
 & \quad\times
 \sqrt{1-2{(p_{V_1}^2+p_{V_2}^2) / M_H^2}
        +{(p_{V_1}^2-p_{V_2}^2)^2 / M_H^4}}
 \qquad .\cr
}
\tag
$$
In this expression we have again again collected the terms in the
numerator that give the on-shell non-running vector boson width, so
again we are not
restricted to use the LO value, but can use the physical value.

Now for the $H\to Z_1\star Z_2\star \to f_1 \bar f_1 f_2 \bar f_2$ we need
to include a symmetry factor of $\half$. This is because when
$f_1 \neq f_2$ we over count the decays of the $Z$'s by counting both
$Z_1 \to f_1 \bar f_1$ ,  $Z_2 \to f_2 \bar f_2$ and
$Z_1 \to f_2 \bar f_2$ ,  $Z_2 \to f_1 \bar f_1$ despite the fact that
these are both the same decay as there is no distinction between $Z_1$
and $Z_2$. When $f_1 = f_2$ then the symmetry factor comes from having
identical particles in the final state.

For off-shell decays of $W$ and $Z$ bosons  we will have
interference between the decays $H\to W\star W\star$ and
$H\to Z\star Z\star$, however these interference effects will only be
large when both vecor bosons are forced off mass shell. Now when both
vector bososns are forced off mass shell the Higgs partial widths will
be exceptionally small and so the interference only contributes a
small term. Hence we ignore this interference in this paper.



The $H\to t\star \bar t\star$ partial width is given by
$$
\eqalign{
\Gamma(H \to t\star \bar t\star) = N_c
{g_{ttH}^2  m_t^4  \over 8 \pi^{3} M_H }
& \int {d(p_t^2) \over \BW_t(p_t^2)}
      {d(p_{\bar t}^2) \over \BW_{\bar t}(p_{\bar t}^2)}
{\Gamma_t(p_t^2) \over p_t^2}
{\Gamma_t(p_{\bar t}^2) \over p_{\bar t}^2} \cr
&\qquad\times
  ((M_H^2-p_t^2-p_{\bar t}^2)(p_t^2+p_{\bar t}^2)/2-2 p_t^2 p_{\bar t}^2)\cr
&\qquad\times
  \sqrt{1+(p_t^2-p_{\bar t}^2)^2/M_H^4-2 (p_t^2+p_{\bar t}^2)/M_H^2}
       ,   \cr
}
\tag
$$
where the tree level running $t$ width is given by
$$
\Gamma_t^{\rm LO} (p_t^2)={g_W^2 \Gamma_W\over 8 \pi^2 M_W m_t }
           \int_0^{p^2_t} {d(p_W^2) \over \BW_W(p_W^2)}
{( p_t^4-2 p_W^4+p_W^2 p_t^2)(1 - p_W^2/p_t^2)}
.\tag
$$
Notice that this integral can be done analytically; however the form
is not particularly illuminating and so we do not give the result
here, although we do use the analytic result in all results.
As the $t$ quark decays via a massive $W$ boson this introduces a
second scale into the problem. This means that unlike the case for
$W$ and $Z$ where we know that for massless decays $m\Gamma \sim p^2$
there is no such simple relationship for $\Gamma_t$. In this paper we
use
$$
\Gamma_t(p^2)={\Gamma_t(m_t^2)\over \Gamma_t^{\rm LO}(m_t^2)}
	          \Gamma_t^{\rm LO}(p^2) \qquad .\tag
$$

\head{Numerical results}
\topinsert
\centerline{\epsfxsize=6 in \epsfbox[55 40 580 405]{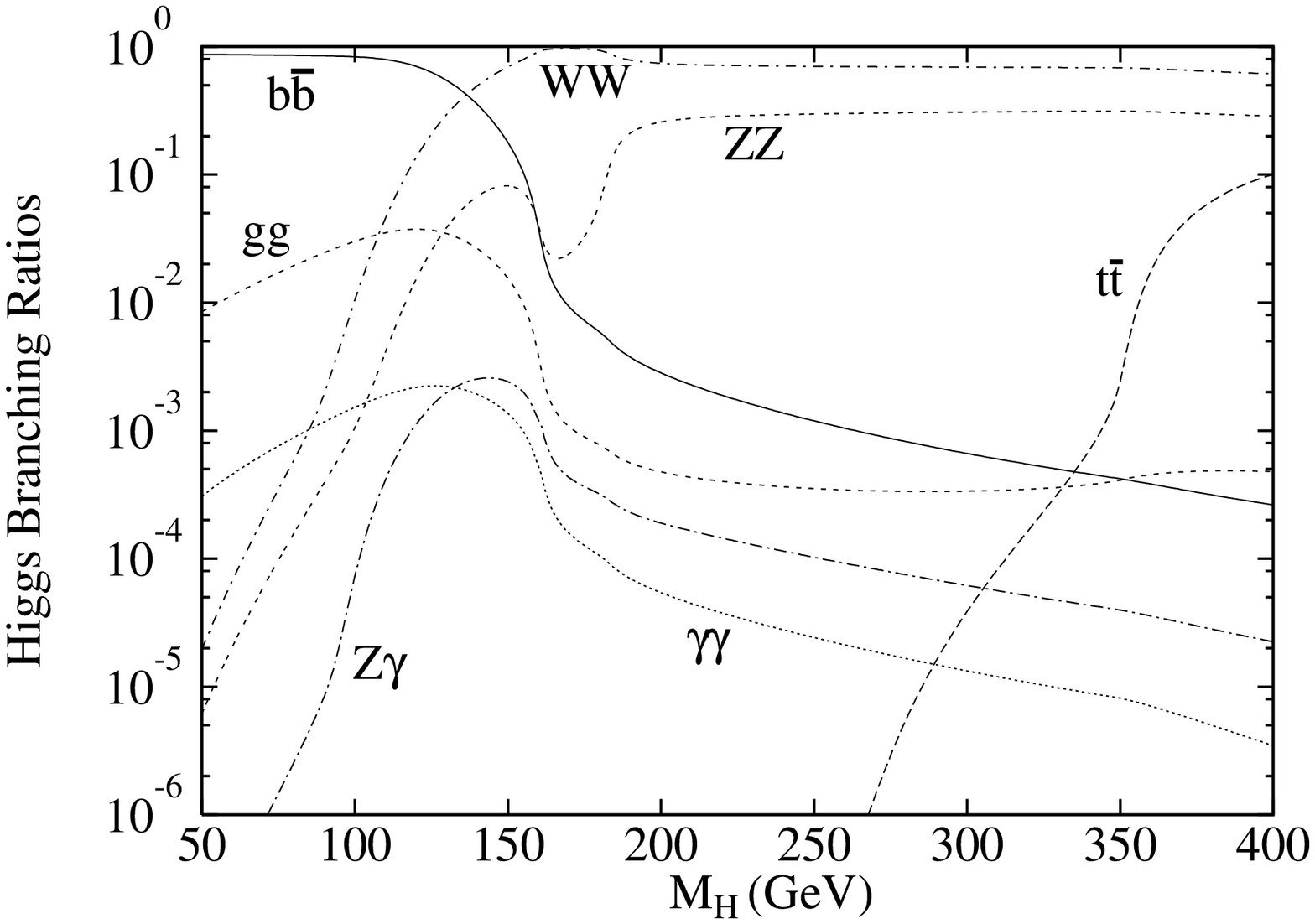}}
\fig{hbr}{Higgs branching ratios.}
\endinsert
In this section we give the numerical results that we obtain for the
formulas in the previous section. In order to present branching ratios
of the Higgs boson we require all partial widths of the Higgs; as we
have only calculated a subset in the previous section we use the
remainder from \ref{Zoltan}. To begin with we show the Higgs branching
ratio as a function of the Higgs mass in \rfig{hbr}. For this we use,
$$\vcenter{\openup1\jot
\halign{\hfil$#$&${}#$\hfil&\hskip 2cm\hfil$#$&${}#$\hfil\cr
m_t&=175\GeV &  \Gamma_t&=\Gamma_t^{\rm LO}= 1.53 \GeV\cr
M_Z&=91.187 \GeV & \Gamma_Z&=2.490 \GeV \cr
M_W&=80.22 \GeV & \Gamma_W&=2.08 \GeV \qquad .\cr
}} \tag
$$
It is clear that for Higgs masses below $2M_W$ and
$2M_Z$ that the branching ratios for $H\to WW\star$ and $H\to ZZ\star$
are still significant. The $H\to Z\star\gamma$ is only significantly
different to the decay to a stable $Z$ for Higgs masses below
$100\GeV$ where the branching ratio is always less than $10^{-4}$ and
so is not likely to be of experimental interest. As the $t$ quark has
turned out to be relatively heavy the Higgs branching ratio is always
dominated by the $WW$ and $ZZ$ decays. This, and that the $t$ width is
relatively narrow, means that when a $t$
quark is forced off shell the $t^*t$ branching ratio is always
smaller than $10^{-3}$ and so it is likely to
be exceptionally hard to experimentally measure.

If we now look at the dependence of these branching ratios on the
width of massive particles, clearly the $Z$ width is already known
very accurately from LEP, and so Higgs decays will not improve the
accuracy of this. The $W$ width is currently measured at hadron
colliders using two different methods. In the first the ratio of
dilepton $Z$ events is compared to single lepton + missing transverse
energy $W$ events. We have
$$
{\sigma(pp \to W \to l\nu) \over \sigma(pp \to Z \to ll) }
= {\sigma(pp \to W) \over \sigma(pp \to Z ) }
 {{\rm Br}(W \to l \nu ) \over {\rm Br}(Z \to l l )}
= {\sigma(pp \to W) \over \sigma(pp \to Z ) }
  {\Gamma_Z \over \Gamma_W }
 {\Gamma(W \to l \nu ) \over \Gamma(Z \to l l )}
.\tag
$$
Now ${\sigma(pp \to W) \over \sigma(pp \to Z ) }$ and
${\Gamma(W \to l \nu ) \over \Gamma(Z \to l l )}$ can be well predicted
within perturbation theory; $\Gamma_Z$ is accurately measured at LEP,
and so this gives a measurement of $\Gamma_W$. Of course this assumes
that ${\sigma(pp \to W) \over \sigma(pp \to Z ) }$ and
${\Gamma(W \to l \nu ) \over \Gamma(Z \to l l )}$ can be accurately
predicted, which is in turn based upon assumptions like the Standard
Model being correct.

A more direct method is to look at the shape of the transverse mass
distribution of $W$ events. If the transverse mass is above the $W$
mass then the decaying $W$ is forced above mass shell; whereas if the
transverse mass is less than the $W$ mass the dominant cross-section
comes from on-shell $W$ decays. Hence, just as with Higgs decays to
virtual $W$'s, the tail of the transverse mass distribution is
sensitive to the $W$ width. Thus a measurement of the shape of this
tail gives a direct measurement of the $W$ width. However the
experimental errors that arise from this more direct method are far
larger than the indirect first method. CDF finds
\refto{indirectW,directW}
$$
\eqalignno{
\Gamma_W^{\rm indirect} &= 2.063 \pm 0.061 {\rm (stat.)}
                                \pm 0.060 {\rm (sys.)} &()\cr
\Gamma_W^{\rm direct} &= 2.11 \pm 0.28 {\rm (stat.)}
                                \pm 0.16 {\rm (sys.)}\qquad , &()\cr
}$$
and so the $W$ width is not particularly accuracy measured, especially
in direct channels.

The $t$ quark width is at the moment totally unmeasured, and there
currently seems little prospect of measuring it at future colliders.

As such we will concentrate on what we can learn
about $W$ and $t$ widths. As we have already mentioned in order to be
sensitive to a massive particle width we need to be beneath the
threshold to produce that particle, as such we shall choose 2
particular Higgs masses to study the $W$ width and $t$ width. For the
$W$ case we shall consider $M_H=150\GeV$. The $WW$
branching ratio of the Higgs is significant for Higgs masses above
$\gsim 110\GeV$, so the Higgs branching ratio will have a similar
sensitivity for all Higgs values between $110 \GeV \lsim M_H < 2 M_W$.
For the $t$ quark case we will consider $M_H=350\GeV\ (=2m_t)$, this
is the largest value of $M_H$ that shows a significant dependence on
$\Gamma_t$, yet the branching ratio is only $0.24\%$; for smaller
values of $M_H$ where the Higgs branching ratio still displays
dependence on $\Gamma_t$ the $H\to tt\star$ branching ratio drops
rapidly. Also as $M_H=2m_t$ is within the threshold region for the
$tt$ decay $\Gamma(H\to t\star t\star) \not\sim \Gamma_t$, and we have
a less simple relationship.

\topinsert
\centerline{
 \epsfxsize=6.5in\epsfbox[55 40 580 405]{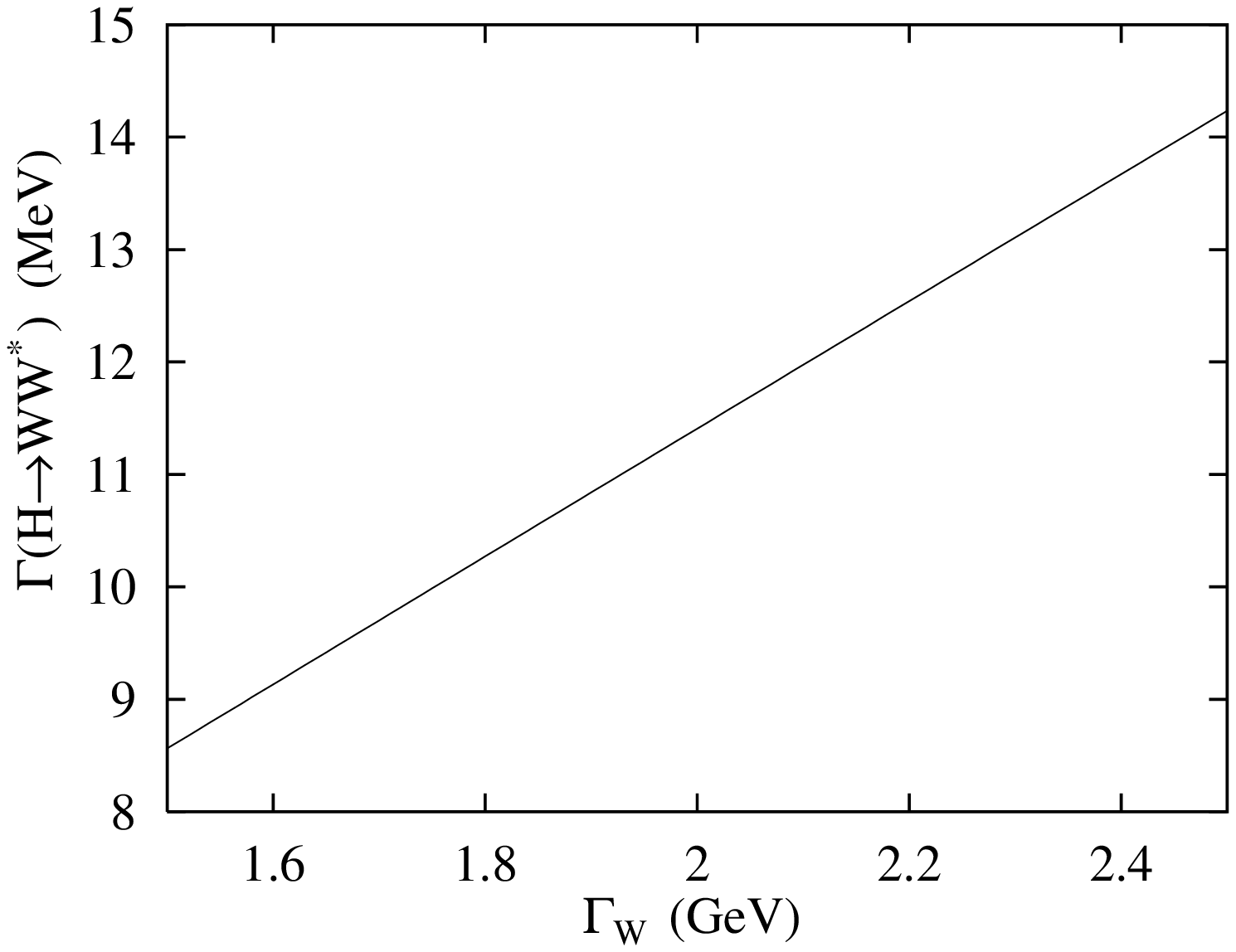}
}
\fig{HWwidth}{The partial width $\Gamma(H\to WW^*)$ as a function of
$\Gamma_W$ for the case $M_H=150\GeV$.}
\endinsert

Considering first the case where $M_H=150\GeV$, in \rfig{HWwidth} we
show the partial width $\Gamma (H\to W\star W\star)$ as a function of
the on-shell $W$ width.
The dependence $\Gamma(H\to W^* W) \sim \Gamma(W)$ is clear to see,
and so a measurement of $\Gamma(H\to W^* W)$ to say 10\% accuracy
gives a measurement of $\Gamma_W$ to the same accuracy.

\topinsert
\centerline{
 \epsfxsize=6.5in\epsfbox[55 40 580 405]{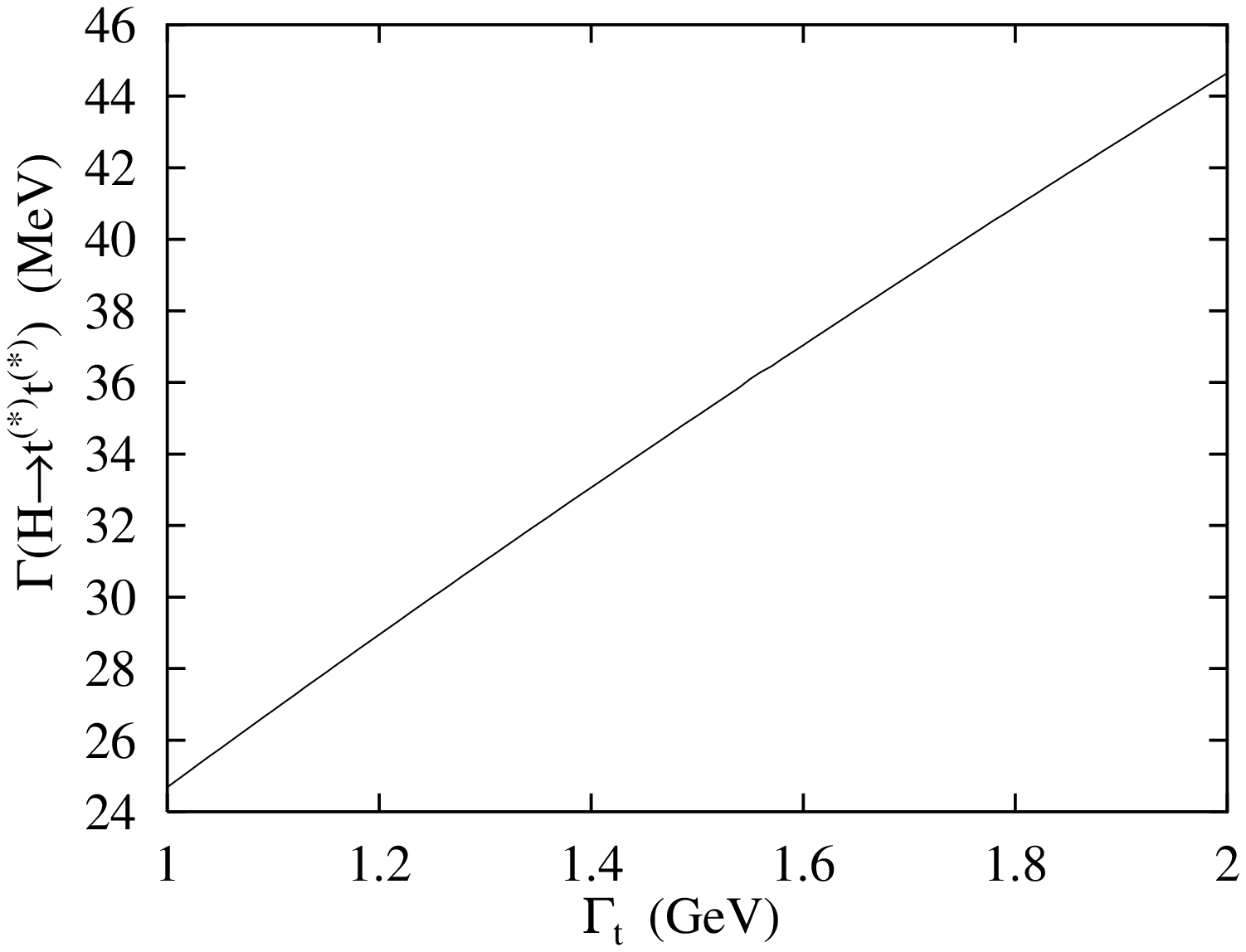}
}
\fig{Htwidth}{The partial width $\Gamma(H\to tt^*)$ as a function of
$\Gamma_t$ for the case $M_H=350\GeV$.}
\endinsert

Moving on to the case where $M_H=350\GeV$ and $m_t=175\GeV$, in
\rfig{Htwidth} we show the width $\Gamma(H\to t\star t\star)$.
$M_H=350\GeV$ is exactly the threshold for the Higgs to decay to
on-mass-shell $t$ quarks; and so for larger Higgs masses
$\Gamma(H\to t\star t\star)$ is independent of the $t$ width,
$\Gamma_t$, whereas for lower Higgs masses we expect,
$\Gamma(H\to t\star t\star) \sim \Gamma_t$. For this Higgs mass
between the two extremes we see a slightly reduced sensitivity
to the $t$ quark width.

\head{Experimental concerns}

At first sight it appears that we are trading a direct measurement
of $\Gamma_W$ for a measurement of $\Gamma(H\to W\star W\star)$, from
which we indirectly extract $\Gamma_W$. It is not immediately clear why
$\Gamma(H\to W\star W\star)$ should be any easier to measure than
$\Gamma_W$. However $\Gamma(H\to W\star W\star)$ is a partial width
whereas $\Gamma_W$ is a total width, and ratios of Higgs decay partial
widths can be measured directly from Higgs branching ratios. For
example we can measure the ratio of say $\Gamma(H\to W^* W)$ to
$\Gamma(H\to Z^* Z)$ by measuring the ratio of Higgs decays to $WW^*$
to Higgs decays to $ZZ^*$. $\Gamma(H\to Z^* Z)$ can be
accurately predicted and so we get a measurement of $\Gamma(H\to W^* W)$.
Hence this method of measuring the $W$ width is, like the CDF
method, an indirect measurement, in which we need an accurate
theoretical prediction of a quantity in order to be able to extract
the $W$ width. We
expect that comparing $H\to WW$ to $H\to ZZ$ to be more accurate than
comparing say $H\to WW$ to $H \to \rm jets$, or ${\rm Br}(H \to WW)$, as
the dominant partial width of the Higgs is for $H\to b\bar b$. The width
$\Gamma(H\to b\bar b)$ has a very
different structure to $H\to VV$, and so for example the radiative
corrections will be somewhat different. Recall that
$\Gamma(H\to b\bar b)$ is decreased by a factor of 2 from leading order
to next--to--leading order, largely due to the running of the $b$ quark
mass which affects the $Hbb$ coupling. Of course it may not be
practical to accurately measure the decay $H \to Z Z^*$ as the branching
ratio for $H\to Z Z^*$ is typically an order of magnitude smaller than
$H\to WW^*$ for  $110\GeV \lsim H_H < 2M_W$.
While we may have enough events to experimentally measure $H\to WW^*$
we will have far fewer $H\to ZZ^*$ events and so may not be able to
measure the rate of these events accurately.

We now turn to ask how we will observe these Higgs events
experimentally. We expect that if the Higgs boson has mass of interest
for measuring massive particle widths, \ie
$110 \GeV \lsim M_H < 2 M_W$ and $M_H\sim 350\GeV$, will first be
observed at hadron colliders. However when a Higgs is produced at a
hadron collider it is usually produced in a messy environment, and
this makes it hard to measure the specific properties of the Higgs.
Typically one hopes just to be able to detect the Higgs in a
particular decay channel. As such there seems few prospects to
measure Higgs branching ratios at future hadron colliders.

On the other hand future high energy $e^+e^-$ and $\gamma\gamma$
colliders offer the opportunity of observing the Higgs in a far cleaner
environment; where the Higgs is produced either with no other
observable particles, or in a relatively simple event. For example
in the processes
$$
\eqalignno{
e^+ e^- &\to ZH &()\cr
e^+ e^- &\to e^+ e^- ZZ \to e^+ e^- H \qquad ,&()\cr
}
$$
the Higgs can be fairly cleanly identified by looking at the mass
that recoils against either the $Z$ or the $e^+ e^-$
pair\footnote{$^\dagger$}{In the case where we have
$e^+ e^- \to ZH \to ZZZ$ notice that we can typically identify which
$Z$ does not come from the Higgs decay, as the mass that recoils
against this $Z$ is equal to $M_H$ which is typically not true for the
$Z$ bosons from the Higgs decay.}.
This would peak very strongly on the Higgs mass
if it were not for initial state radiation off the $e^+e^-$ pair; in
practice this radiation smears the recoil mass somewhat, however there
is still a sharp peak at the Higgs mass, which can be used as a tag
for Higgs events \refto{red}.
For example at a NLC with $\sqrt s = 300\GeV$ and
$\int {\cal L} = 10\, {\rm fb}^{-1} {\rm year}^{-1}$ then for
$M_H=150\GeV$ we expect
${\cal O}(1000)$  $ZH$ Bjorken events per year \refto{red}.

Having identified these Higgs events one
can easily
look at the decay products of the Higgs. If we are interested in
measuring the $W$ width then we want to be able to distinguish decays
of Higgs to $WW$ from other decays of Higgs, in particular Higgs
decays to $ZZ$. If both $W$'s decay hadronically then we are unlikely
to be able to distinguish the $W$ decays from $Z$ decays; however if
at least one $W$ decays leptonically then we can distinguish the $W$
which decays into a single observed lepton, and the $Z$ which decays
into a pair of leptons. So if $M_H=150\GeV$ and we produce 1000 tagged
Higgs we would expect to have ${\cal O}(250)$ $WW^*$ events tagged by
a single $W$ decaying to an isolated lepton + missing energy; on the
other hand we would only have ${\cal O}(20)$ $ZZ^*$ events tagged for
a single $Z$ decaying leptonically. Thus with only ${\cal O}(1000)$
Higgs events we will have enough tagged $WW^*$ events that the
statistical error on the event rate is small, whereas the statistical
error on the $ZZ^*$ event rate is still relatively large. This means
that with only ${\cal O}(1000)$ Higgs events we do better by
measuring the $H\to WW^*$ branching ratio and extracting
$\Gamma(H\to WW^*)$ from that.

However if we measure specific decay products of the $W$ and $Z$
bosons then it is only that specific width that enters in the
numerator of the Higgs partial width. For example if we observe the
decay
$$
\eqalign{
H \to &WW^* \to l\nu \cr
\noalign{\vskip -9pt}
&\hskip 4pt\barrow \hbox{\rm jets} \cr
}\qquad ,\tag
$$
then the Higgs width for this process is proportional to the width
$\Gamma(W \to l\nu)$ rather than the total $W$ width; of course if we
observe
$$
\eqalign{
H \to &W^* W \to l\nu \cr
\noalign{\vskip -9pt}
&\hskip 4pt\barrow \hbox{\rm jets} \cr
}\qquad ,\tag
$$
then this is proportional to $\Gamma(W \to {\rm jets})$. So by measuring
different off-shell decays of the Higgs we get a measurement of
different partial width of that massive particle. Of course the ratio
of these two rates is just the ratio of off-mass-shell $W$
branching ratios, and we expect these to be very similar to the
on-mass-shell $W$ branching ratios which we expect will be accurately
measured at LEPI$\!$I.

Notice that
typically for a Higgs mass in the range $110\GeV \lsim M_H < 2 M_W$
only a single $W$ is forced off mass shell, the other $W$ is typically
produced on mass shell and so shows no dependence on the $W$ width.

\head{Theoretical and experimental accuracy}

We should ask how accurate are our theoretical predictions of these
partial widths of the Higgs.

Rewriting the numerator of $\Gamma(H\to Z\star\gamma)$ and
$\Gamma(H\to V\star V\star)$ as $\Gamma_V$
includes all the radiative corrections associated with the decay of
$V$. This means that we are only vulnerable to radiative corrections to
the production of vector bosons, and also to interference between the
the decay products of the 2 vector bosons. The former corrections are
only electroweak in nature, and so we expect them to be
${\cal O} (\alpha_{\rm em})$ and so small; the latter we expect to be
suppressed as the two vector bosons will decay at different
time scales\refto{Khoze}. As a result we expect our results to be
accurate to a few percent.

For the case $\Gamma(H\to t\star \bar t \star)$, again rewriting the
$t$ width in the numerator effectively includes all the radiative
corrections associated with the decay of the $t$ quark, however in
this case the radiative corrections to the production of a $t$ $\bar
t$ pair are QCD in nature; so we expect the radiative corrections to
be ${\cal O}(\alpha_S)$ or 10\% or so. This means that the current
calculation could not be used to make accurate measurements of the $t$
quark width. However the dependence on the $t$ quark width that our
calculation shows we would expect to be reproduced in a more accurate
calculation. As such our results should be taken as representative of
the kind of accuracy that one can expect in measuring the $t$ quark
width if one had a more accurate calculation including the QCD
corrections.

If there is physics beyond the Standard Model then we should worry
that the decay of a Higgs to a virtual massive particle is not that
predicted by the Standard Model. Indeed such physics beyond the
Standard Model is likely to show up in the production of the massive
particle, rather than its propagation and decay which are already
measured at current colliders. Now if the production of the massive
particle differs from the Standard Model prediction then we lose all
ability to measure the massive particle width until that physics is
understood to the level to which we wish to measure the massive
particle width.  However if we measure the $W$ width by comparing the
rate of $H\to W^*W$ decays to $H\to Z^* Z$ decays then as we expect
the $W$ and $Z$ masses to be generated by the same symmetry breaking
mechanism we expect that the ratio of their couplings to be
independent of the physics beyond the Standard Model, that is
$g_{HWW}/g_{HZZ}=M_W^2 / M_Z^2$.  Thus if the production rate of $H\to
W^* W$ is changed we would expect to change the production rate of
$H\to Z^* Z$ by a similar amount. This means that the ratio of $WW^*$
decays to $ZZ^*$ decays is fairly insensitive to physics beyond the
Standard Model. For other ratios of Higgs decays we are not so lucky
and beyond-the-Standard-Model physics will probably have drastic
effects, making a measurement of massive particle widths impossible,
at least until the additional physics is accurately understood.

\topinsert
\centerline{
 \epsfxsize=6.5in\epsfbox[55 40 580 405]{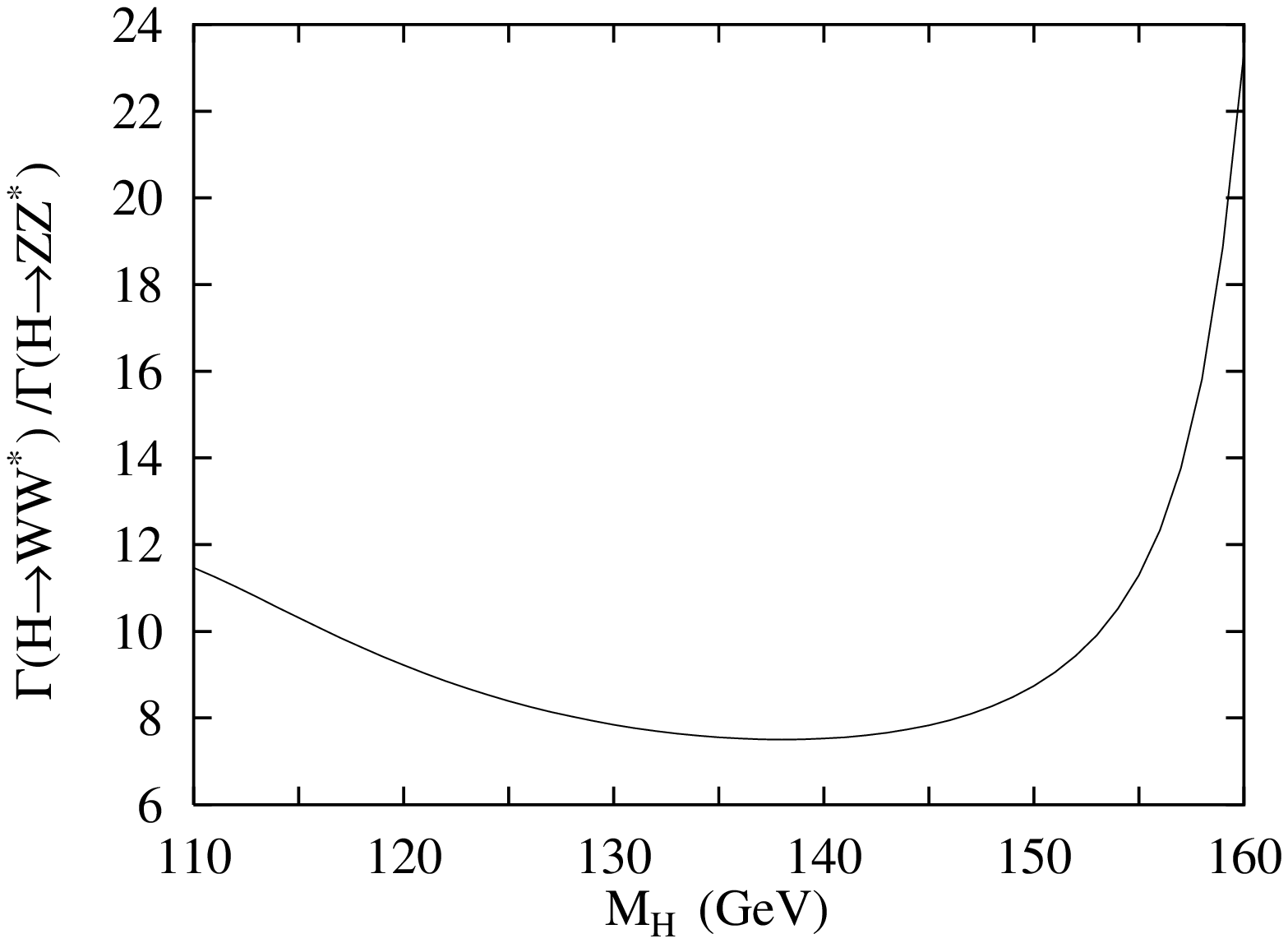}
}
\fig{MHdep}{The ratio of the partial width of $\Gamma(H\to WW^*)$ to
$\Gamma(H\to ZZ^*)$ as a function of $M_H$.}
\endinsert

An experimental difficulty arises if we do not know the exact
Higgs mass, as the theoretical partial widths can often vary rapidly
with the Higgs mass. So for example if the Higgs mass is 1\% higher
than $150\GeV$ the $\Gamma(H\to WW^*)$ partial width is 16\% higher!
So it seems that to measure the $W$ width to 10\% we need to know
the Higgs mass to about $1\GeV$. However we can decrease this
sensitivity on the Higgs mass by comparing different decays of the
Higgs, as we have to do if we are to measure the $H\to W\star W\star$
width. For
example the $H\to Z\star Z\star$ partial width drops rapidly for
lighter Higgs just like the width for $H\to W\star W\star$, so the
ratio of Higgs decays to $W\star W\star$ and $Z\star Z\star$ is far
less sensitive to $M_H$, while still retaining sensitivity to the $W$
width. We show the ratio
$\Gamma(H\to W\star W\star)/ \Gamma(H\to Z\star Z\star)$ as a function
of the Higgs mass in \rfig{MHdep}.

Clearly this ratio has far less dependence on the Higgs mass for Higgs
masses less than $155\GeV$, and so an accurate knowledge of $M_H$ is
not so crucial. However for Higgs masses over $155\GeV$ the width
$\Gamma(H\to W\star W\star)$ grows very rapidly as the $W$'s start to
come on mass shell, this means that even the ratio $\Gamma(H\to W\star
W\star)/ \Gamma(H\to Z\star Z\star)$ is very sensitive to the Higgs
mass and so is not a good means to measure the $W$ width. As
$\Gamma(H\to Z\star Z\star)$ is growing more rapidly with the Higgs
mass than all other partial widths, with the exception of $\Gamma(H\to
W\star W\star)$, all other ratios of the Higgs partial widths will
show far greater dependence on the Higgs mass. So we must conclude
that for Higgs masses $155\GeV \lsim M_H < 2M_W$, although the
$\Gamma(H\to W\star W\star)$ width is still proportional to the the
$W$ width, that measuring it is not a practical method for measuring
the $W$ width, unless the Higgs mass is known exceptionally well.

In the case where we look at measuring the $t$ quark
width the situation is clearly worse. For $M_H=350\GeV$ the
$H\to t\star t\star$ width is clearly growing very rapidly due to the
narrow $t$ quark width, and no other partial width of the Higgs shows
anywhere near as rapid growth. This means that for $M_H=350\GeV$ we
must know the Higgs mass to about $0.3\GeV$ just to measure the $t$
quark width to 10\% accuracy. On the other hand for lower Higgs masses
where the $H\to t\star t\star$ width grows less rapidly (but still
quickly) with the Higgs mass the partial width
$\Gamma(H\to t\star t\star)$ is exceptionally small, also due to the
narrow $t$ quark width, and so will be exceptionally hard to detect
experimentally.

\head{Conclusions}

In this paper we calculate the decays of the Standard Model Higgs
that proceed via a massive intermediate particle. These results differ
from those previously published due to a more careful treatment of the
form of the Breit Wigner propagator for the massive particle.

We calculate the branching ratios of a Standard Model Higgs with
the currently available values for the physical parameters.

We show that the rate for the Higgs to decay via an intermediate
virtual massive particle is regulated by that massive particle's width;
as such this gives a possible method for measuring massive particle
widths if the Higgs happens to have a convenient mass.

We briefly discuss how one would hope to experimentally measure the
partial widths of the Higgs boson at future colliders. We also discuss
how the partial widths are affected by higher order corrections, and
experimentally measured parameters. In particular we discuss how these
partial widths to a virtual particle depend sensitively on the Higgs
mass. Often this dependence is so sensitive that we can imagine that
measurement of these widths will give a potentially accurate measure of
the Higgs mass. However in the current case where we are interested
in measuring these partial widths as a means of measuring massive
particle widths, this great sensitivity on the Higgs mass reduces our
ability to measure the massive particle width. This means that we
either need to know the Higgs mass exceptionally accurately, or form
the ratio of the massive particles partial width with another Higgs
decay which shows a similar dependence on the Higgs mass. Currently
there are two massive particles whose widths are not known
particularly accurately, the $W$ boson and the $t$ quark. For the $W$
boson we can form the ratio
$\Gamma(H\to W\star W\star ) / \Gamma(H\to Z\star Z\star )$ which is
relatively insensitive to the Higgs mass if $H_H\lsim 155\GeV$ yet
still retains its sensitivity to the $W$ width. For the $t$ quark
there is no ratio that removes the strong dependence on the Higgs
mass; this, and that the $H\to t\star t\star$ drops very rapidly
below the $tt$ threshold, means that it is not practical to measure
the $t$ quark width in Higgs decays.

\vfil\eject
\headnn{Acknowledgements}

I would like to thank Dieter Zeppenfeld and Jim Amundson for helpful
conversations on the form of a massive fermion Breit Wigner, Bernd
Kniehl for helful thoughts on the process $H\to Z^* \gamma$, and
Jim Amundson for reading the manuscript.

This research was supported in part by the University of Wisconsin Research
Committee with funds granted by the Wisconsin Alumni Research Foundation, in
part by the U.S.~Department of Energy under Contract No.~DE-AC02-76ER00881,
and in part by the Texas National Research Laboratory Commission under
Grant No.~RGFY93-221.

\references

\refis{indirectW} CDF Collaboration, {\it Phys.Rev.Lett.} {\bf 73}
(1994) 220.

\refis{directW} CDF Collaboration, {\it Phys.Rev.Lett.} {\bf 74}
(1995) 341.

\refis{thesis} D.J.Summers, University of Durham, Ph.D. thesis, unpublished.

\refis{red} A.Djouadi \etal, Proceedings of the $e^+e^-$ Collisions at
$500\GeV$ Workshop -- Munich, Annecy, Hamburg, DESY 92-123A.

\refis{RJNP} A.Grau, G.Pancheri, and R.J.N.Phillips,
{\it Phys.Lett.B} {\bf 251} (1990) 293.

\refis{HHG} A.Barroso, J.Pulido and J.C.Rom\~ao,
{\it Nucl.Phys.} {\bf B267} (1986) 509.\hfill\break
J.C.Rom\~ao and A.Barroso,
{\it Nucl.Phys.} {\bf B272} (1986) 693.

\refis{fermprop} M.Nowakowski, and A.Pilaftsis, {\it Z.Phys.C}
{\bf 60} (1993) 121.

\refis{Zoltan} Z. Kunszt, 'Perspectives on Higgs
Physics', ed. G. Kane, World Scientific Publ. (1992).

\refis{Khoze}  Torbjorn Sjostrand, and Valery A. Khoze,
{\it Z.Phys.C} {\it 62} (1994) 281.

\refis{Cern} Proceedings of the Workshop on Physics at Future Accelerators La
Thuile, ed. J. Mulvey, CERN Yellow Report 87-07 (1987).

\refis{brH} Z.Kunszt and W.J.Stirling, Aachen ECFA Workshop (1990) 428.

\endreferences

\endit